\documentclass[aps,nofootinbib,showpacs]{revtex4}

\usepackage{epsfig}

\begin{document}

\title{Neutrino Mass Inference from SZ Surveys}

\author{Meir Shimon$^1$}
\author{Sharon Sadeh$^2$}
\author{Yoel Rephaeli$^{1,2}$}

\affiliation{$^1$ Center for Astrophysics and Space Sciences, University
of California, San Diego, 9500 Gilman Drive, La Jolla, CA, 92093-0424} 
\affiliation{$^2$ School of Physics and Astronomy, Tel Aviv University, 
Tel Aviv 69978, Israel}

\begin{abstract}

The growth of structure in the universe begins at the time of 
radiation-matter equality, which corresponds to energy scales 
of $\sim 0.4 eV$. All tracers of dark matter evolution are 
expected to be sensitive to neutrino masses on this and smaller 
scales. Here we explore the possibility of using cluster number 
counts and power spectrum obtained from ongoing SZ surveys to 
constrain neutrino masses. Specifically, we forecast the capability 
of ongoing measurements with the PLANCK satellite and the ground-based 
SPT experiment, as well as measurements with the proposed EPIC 
satellite, to set interesting bounds on neutrino masses from their 
respective SZ surveys. We also consider an ACT-like CMB experiment 
that covers only a few hundred ${\rm deg^{2}}$ also to explore the 
tradeoff between the survey area and sensitivity and what effect this 
may have on inferred neutrino masses. We find that for such an 
experiment a shallow survey is preferable over a deep and low-noise 
scanning scheme. The precision with which the total neutrino mass can be 
determined from SZ number counts and power spectrum 
is limited mostly by uncertainties in the basic cosmological parameters, 
the mass function of clusters, and their mean gas mass fraction; all 
these are explicitly accounted for in our statistical Fisher-matrix 
treatment. We find that projected results from the PLANCK SZ survey 
can, in principle, be used to determine the total neutrino mass with 
a ($1\sigma$) uncertainty of $0.28 eV$, if the detection limit of a 
cluster is set at the $5\sigma$ significance level. This is twice as 
large as the limits expected from PLANCK CMB lensing measurements. The 
corresponding limits from the SPT and EPIC surveys are $\sim 0.44 eV$ 
and $\sim 0.12 eV$, respectively. 
Mapping an area of 200 deg$^{2}$, ACT measurements are predicted to attain 
a $1\sigma$ uncertainty of 0.61 eV; expanding the observed area to 4,000 
deg$^{2}$ will decrease the uncertainty to 0.36 eV.

\end{abstract}

\maketitle

\section{Introduction}

The temperature anisotropy and polarization state of the CMB constitute a 
unique window on the physical state of the universe in the era of 
recombination ($z\approx 1100$), $\sim 400,000$ years after the big bang. 
Measurements with the COBE and WMAP satellites, along with a host 
ground-based and balloon-borne experiments, have led to impressively precise 
values of the basic cosmological parameters. Other cosmological probes, 
sensitive to certain combinations of these parameters, supplement CMB data, 
significantly reduce parameter degeneracies, and provide some control of 
systematic uncertainties. A major advantage of the CMB is its well understood 
underlying physics, a fact that justifies its distinction as the most precise 
cosmological probe. This said, it is also well recognized that the diagnostic 
power of the primary CMB anisotropy, whose main features reflect conditions at 
recombination, is rather limited in constraining certain cosmological parameters. 
Measurements with the recently launched PLANCK satellite are 
cosmic-variance-limited up to $\ell \approx 2500$, and are therefore expected 
to reduce this limitation significantly by mapping the primary CMB anisotropy 
down to scales of a few arcminutes.

Neutrino masses are a good example for the need to sensitively map the primary 
CMB anisotropy at high $\ell$. 
Below their Jeans scale, massive neutrinos free-stream and act as a hot 
dark matter (HDM) component, reducing the growth of the large scale structure 
(LSS) as compared with the evolution of structure in a pure cold dark matter (CDM) 
model. This characteristic neutrino signature provides a diagnostic measure 
that can be exploited to either determine or constrain their masses. 
Current cosmological constraints on neutrino masses come from the CMB 
temperature anisotropy e.g. [1,2], weak gravitational lensing [3-6], 
as well as galaxy [7-11] and Ly$\alpha$ surveys [12-15]. 
WMAP7 data already set an upper limit (at the 95\% confidence level) of 0.58 eV 
on total neutrino mass [2]. This limit, which comes mainly from CMB constraints 
on the {\it early} integrated Sachs Wolfe (ISW) effect ($l<200$), is related to 
the fact that the neutrino temperature at radiation-matter equality is 
$\approx 0.4$ eV (in equivalent energy units), and that there are three active 
neutrino species. The ISW effect is associated with the kinematics of neutrinos 
at the time period between radiation-matter equality and recombination. This is 
an indirect probe of neutrino kinematics at, and prior to, last scattering. To 
improve on this bound, probes of neutrino kinematics at lower redshifts are 
required, where neutrino temperatures are lower (it can certainly be the case 
that a massive neutrino which is nonrelativistic today, was actually 
relativistic, i.e. freely streamed, at much higher redshifts). This may be 
achieved, for example, by employing lensing extraction of the CMB, e.g. [16-18].

The relatively strong bounds deduced from CMB data on the {\it total} neutrino 
mass should be contrasted with the mass that would have provided closure density 
(of a hypothetical neutrino-dominated) universe, $\approx 45$ eV, and to the 
actually measured value of the matter density parameter, $\Omega_{cdm}\approx 
0.3$ (equivalently, $\sim 5 eV$ per neutrino species if the entire dark matter 
was made up of neutrinos). From this perspective, the total neutrino mass is 
bounded to be lower than $4\%$ ($2\sigma$) of all dark matter (DM). As a 
tracer of LSS, gravitational lensing of the CMB is a sensitive probe 
of any cosmological parameters that affect the formation or growth-rate of the 
perturbed background matter density. 

Galaxy clusters are excellent probes of structure formation, and therefore can 
be used, in principle, for indirectly setting limits on neutrino masses through 
their dynamics and impact on structure formation. For this purpose, it has been 
suggested [19] that cluster number counts and cluster correlations can be used 
to set useful upper bounds on neutrino masses. These probes are sensitive to the 
evolution of structure down to the typical correlation scale of galaxy clusters, 
a few tens of Mpc, and are therefore sensitive to sub-eV neutrino masses. Lensing 
of the CMB with planned good sensitivity and high resolution experiments, e.g. 
[20,21] could do even better, perhaps even constrain the total neutrino mass, 
$M_{\nu}$, at the 0.05 eV level. It is interesting to note that this level overlaps 
with the lower mass limit from neutrino oscillations, e.g. [22-24]. This implies 
that, to the level of confidence we have in these forecasts, in the not too distant 
future, neutrino masses and mass hierarchy will be resolved, if next generation 
CMB satellites (e.g. EPIC [20]) will indeed be launched. However, other probes are 
certainly of interest even if their respective precision levels cannot compete with 
the CMB lensing probe, since they can provide much needed consistency checks for 
systematics and parameter degeneracy breaking.

We have explored the possibility of determining neutrino masses from survey 
measurements of the SZ effect (Comptonization of the CMB by hot gas in clusters, 
e.g. [25-27]) by the currently operational ground-based SPT project, PLANCK 
satellite, and the proposed EPIC satellite mission. 
We also explored various scanning schemes for the ground-based ACT to gain some 
insight on how to optimize SZ surveys for neutrino mass inference. 
The basic motivation for doing so is provided by the fact that the SZ effect 
is a sensitive tracer of the evolution of the LSS. However, in contrast to the 
linear physics characterizing much of the evolution of the primary CMB 
anisotropy, the SZ effect is more model-dependent (e.g., mass-temperature 
scaling relation, gas mass fraction and spatial profile). 
A substantial source of uncertainty is the distribution of the entire cluster 
population in mass-redshift space, i.e. the mass function. This important 
function, whose specific shape and normalization reflect the details of the 
growth of density fluctuations, and the non-linear hierarchical collapse and 
mergers of sub-structures, can be best studied by state-of-the-art large-volume 
hydrodynamical cosmological simulations. Although very advanced, currently 
available computer codes predict a range of mass functions. At present, this 
mass function indeterminacy largely sets the precision limit of forecasting 
the total neutrino mass from cluster SZ number counts and power spectrum.

In this paper we discuss constraints on neutrino masses that can be 
obtained from SZ surveys alone. In particular, we determine how likely it 
will be for PLANCK, SPT and EPIC to constrain neutrino masses by using the 
sensitivity of cluster number counts and SZ power spectrum to neutrino masses. 
For the first time in this context, attention is given to the robustness 
of the results to the assumed fiducial neutrino mass. One of our main 
objectives is to check to what extent the upper limits on neutrino 
masses derived from SZ surveys are robust to modeling uncertainties and 
other systematics.

The paper is organized as follows. In section II we discuss the impact 
of massive neutrinos on the LSS with special emphasis on its 
manifestation in the transfer function and cluster mass function. 
Cluster number counts, correlation, and induced CMB anisotropy are 
described in section III. We lay out the methodology of our estimates 
for expected number counts and SZ power as a function of neutrino mass 
and address issues such as detection threshold, foregrounds, etc. along 
with a description of the Fisher matrix analysis used in this work in 
section IV. Our main results are presented in section V and further 
discussed in section VI.

\section{Neutrino Impact on Growth of the LSS}

At a redshift $3200$, the CMB and matter energy densities, together with 
the neutrino contribution, set the time of matter-radiation equality, 
$t_{eq}$. This transition preceded recombination, but not by much; thus, 
the universe was not purely matter-dominated at recombination. Consequently, 
gravitational potential wells could decay and induce the early ISW 
effect, which boosts the angular power spectrum of the CMB temperature 
anisotropy on multipoles that correspond to the horizon scale at 
recombination, i.e. $l\sim 100-200$. In other words, $a_{rec}/a_{eq}$, 
the ratio of the scale factors at recombination and matter-radiation 
equality is a measure of the linear growth of structure during the 
period between radiation-matter equality and recombination. This ratio 
clearly depends on neutrino masses. The fraction of total energy in the 
form of radiation (which would include neutrinos, if the total neutrino 
mass is below $\sim 1 eV$) at recombination determines the amplitude of 
the early ISW effect. 

More stringent constraints on neutrino masses could, in principle, be obtained 
from the SZ effect. A massive neutrino that is nonrelativistic today 
[$m_{\nu}c^{2}>kT_{\nu}(0)$], was relativistic at higher redshifts 
[$m_{\nu}c^{2}<kT_{\nu}(z)$], and therefore could not easily cluster then, 
causing a relative suppression of structure formation compared to what would be 
expected had this species been non-relativistic for the entire history of structure 
formation ($t>t_{eq}$). In principle, precise measurements of the LSS - for 
example, cluster number counts - can be used to place constraints on neutrino 
masses.

The evolution of structure in the matter-dominated era is described in terms of 
the matter power spectrum,
\begin{eqnarray}
P_{m}(k,z)=Ak^{n}T^{2}(k,z),
\end{eqnarray}
with the primordial density fluctuation spectrum, $Ak^{n}$, where 
$A$ is an overall normalization, $n$ is the tilt of the power spectrum, 
and the transfer function, $T(k,z)=T(M_{\nu};k,z)$. An important quantity 
gauging the amplitude of the processed power spectrum observed today is 
the mass variance parameter on a scale of $8Mpc\ h^{-1}$, 
\begin{eqnarray}
\sigma_{8}=\int_{0}^{\infty}P_{m}(k,z)W^2(kR)k^2\frac{dk}{2\pi^{2}},
\end{eqnarray}
where $W(kR)$ is a window function, and $R = 8Mpc\ h^{-1}$. The latter quantity 
incorporates the physics of neutrino damping; thus, $\sigma_{8}$ is a function 
of not only $A$, $n$, but also of neutrino masses (via the transfer function), 
and indeed any other cosmological parameter which affects structure evolution on 
scales of few tens $Mpc$ and lower. Because these scales are comparable to typical 
scales of diffusion damping of density fluctuations by neutrinos with small masses, 
we expect that $M_{\nu}$ and $\sigma_{8}$ will be anti-correlated. Since the SZ 
signature is a strong function of $\sigma_{8}$, it is expected to be sensitive also 
to $M_{\nu}$.

\subsection{The Transfer Function}

Quantitative description of the effects of neutrinos on the evolution of 
the LSS begins with a modification of the density fluctuation spectrum as 
described by the transfer function. Whereas the standard CDM transfer 
function is separable in the perturbation wave-number $k$ and redshift $z$
\begin{eqnarray}
T(k,z)=D(z)\tilde{T}(k)
\end{eqnarray}
where $D(z)$ is the linear growth factor and, e.g. [28],
\begin{eqnarray}
\tilde{T}(k)=\frac{\ln(1+2.34q)}{2.34q}
[1+3.89q+(16.1q)^{2}+(5.46q)^{3}+(6.71q)^{4}]^{-1/4}
\end{eqnarray}
and $q\equiv k/(\Omega_{m}h^{2}) Mpc^{-1}$, a mixed DM model 
generally cannot be cast in such a simple form. This is due to the 
redshift-dependence of the free-streaming scale, $k_{FS}$, which is determined 
by both neutrino mass and the redshift-dependent neutrino temperature, 
$T_{\nu}(z)$. A reasonable first approximation to the transfer function in 
the presence of neutrinos, is e.g., [29]
\begin{eqnarray}
T^{2}(k,z)&\rightarrow&D^{2}(z)\tilde{T}^{2}(k);\ \ \ k<k_{nr}\nonumber\\
T^{2}(k,z)&\rightarrow&D(z)^{2-\frac{6}{5}f_{\nu}}
(1-8f_{\nu})\tilde{T}^{2}(k);\ k>k_{nr}.
\end{eqnarray}
In the following this approximation is merely used in justifying order of 
magnitude sensitivity of the SZ power spectrum to changing the fiducial neutrino mass;
in our numerical calculations we employ a publically available 
code [30] to {\it precisely} determine $T(k,z)$.
The effect of including nonlinear terms in $f_{\nu}$ on the matter power 
spectrum was investigated in several recent works, e.g. [31-33].
Indeed, this higher order effect modifies the matter power spectrum 
on scales of few Mpc and smaller, and in principle should be accounted for 
when data from future SZ surveys is available. In the current work 
this effect is ignored for simplicity. We can justify doing so by noting 
that the extra $O(f_{\nu}^{2})$ term will only increase the sensitivity 
of $\sigma_{8}$, and consequently also of the SZ effect, to the neutrino 
mass and thereby will increase the sensitivity of SZ-based 
cluster number counts and power spectrum to neutrino mass. 
In this regard the present analysis 
represents a conservative assessment of the capacity of cluster number 
counts and cluster correlations to constrain neutrino mass. Indeed, the latter 
studies [32,33] show that the factor $1-8f_{\nu}$ should be replaced by 
$\approx(1-9.8f_{\nu})$ for $k\gg k_{nr}$. It is also 
clear that the $f_{\nu}^{2}$ corrections to the matter power spectra are 
important when $f_{\nu}$ is sufficiently large. Indeed, neutrino masses twice 
or three times as large as the current 95\% confidence upper limit on neutrino 
mass from WMAP (0.58 eV) were considered [31-33]. For the range of neutrino 
masses considered in this work, $M_{\nu}\leq 0.6$ eV, the nonlinear neutrino 
effects are smaller.

\subsection{Cluster Mass Function}

The basic quantity which describes the 
number density of clusters as a function of their mass and 
redshift - basic properties by which clusters are identified - 
is the mass function. As will be shown later, the total neutrino mass can be 
derived from comparison of the observed number of clusters at a given redshift-bin 
to the number predicted from the mass function, $\frac{dn(M;z)}{dM}$. 
The latter function is defined in terms of the differential cluster number 
in a redshift interval,
\begin{eqnarray}
dN(M,z)=f_{sky}\frac{dn(M,z)}{dM}dVdM ,
\end{eqnarray}
where $dV$ is a volume element and $f_{sky}$ is the observed sky fraction. 
The total number in a given interval $\Delta z$ 
around $z_{i}$ is
\begin{eqnarray}
\Delta N(z_{i})=f_{sky}\Delta z_{i}\frac{dV(z_{i})}{dz}\int\frac{dn(M,z_{i})}{dM}dM.
\end{eqnarray}
As noted earlier, the currently most optimal determination of the mass 
function is from extensive cosmological simulations. Here we adopt the mass 
function derived recently from a large set of collisionless cosmological 
simulations in $\Lambda$CDM cosmological model [34]. 
Expressing the mass function in the familiar form, 
\begin{eqnarray}
\frac{dn}{dM}=f(\sigma)\frac{\rho_{m}}{M}\frac{d\ln(\sigma^{-1})}{dM} ,
\end{eqnarray}
these authors derived a fit of the form 
\begin{eqnarray}
f(\sigma)=A\left[\left(1+\frac{\sigma}{b}\right)^{-a}\right]
e^{-\frac{c}{\sigma^{2}}} ,
\end{eqnarray}
where the parameters $A$, $a$, $b$ and $c$ were best-fit to the results of their 
simulations. The mass function does not have a universal form, a fact 
that becomes apparent by the deduced dependence of these fit parameters on 
both redshift and the overdensity at virialization, $\Delta_{v}$.

\section{SZ Power Spectrum}

The SZ effect is a unique probe of cosmological parameters and cluster 
properties; its statistical diagnostic value is reflected through number 
counts and the power spectrum of the CMB anisotropy it induces. 
The dependence of cluster number counts and cluster correlations 
on the total neutrino mass, and an assessment of the feasibility 
of actually determining it from their measurements, are our main 
objectives in this work. To this end, we carry out a precise 
calculation of the SZ signature imprinted by a cluster of mass $M$ 
at redshift $z$. We then calculate the integrated 
statistical pattern of the SZ-induced CMB anisotropy projected on the 
sky, and the corresponding cluster number counts. Obviously, the 
ultimate practical issue is how well can clusters be detected by a 
specific CMB experiment (characterized by both detector noise and 
angular resolution) in the presence of the primary CMB anisotropy and 
the emission of foreground (point) sources. 

The thermal component of the SZ effect induced by a cluster with optical depth 
$\tau\equiv\int n_{e}\sigma_{T}dl$ and gas temperature $T_{e}$ induces a 
small level of fractional change in the CMB temperature [35,36]
\begin{eqnarray}
\frac{\Delta T}{T}=y\mathcal{F}(x)
\end{eqnarray}
where $y\equiv\tau\Theta_{e}$ is the Comptonization parameter, 
$\Theta_{e}\equiv kT_{e}/(m_{e}c^{2})$, and $k$, $m_{e}$, and $c$ 
are the Boltzmann constant, electron mass and 
speed of light, respectively. $n_{e}$ and $\sigma_{T}$ are the gas number 
density and Thomson cross section, respectively, and the integration over 
the photon pathlength is along the line of sight.
In the classical limit the spectral function reads
\begin{eqnarray}
\mathcal{F}(x)=x\coth(x/2)-4
\end{eqnarray}
where $x\equiv h\nu/(kT)$ is the dimensionless CMB frequency, 
and $h$ and $T$ denote the Planck constant and CMB temperature, 
respectively.
Note that in this work we ignore relativistic corrections [37, 38], 
whose weighted impact over the full temperature range of clusters is small. 
Since $n_{e}$ is non-uniform and has a typical scale, the cluster core radius $r_{c}$ 
(as well as its temperature $T_{e}$ which we here assume uniform $T_{e}(r)=T_{0}$ 
for simplicity), the projected 2D profile of the $y$-parameter on the 
sky, inherits this typical angular scale, that determines 
together with other parameters (as we discuss below) the SZ flux associated 
with the cluster and thereby the likelihood of its detection 
with a given CMB survey. The 2D Fourier transform of the projected 3D 
Comptonization parameter generated by an individual cluster is, e.g. [39]
\begin{equation}
\tilde{\xi}_{\ell}
=\frac{4\pi r_c}{l_c^2}\int_0^{\infty}r^{2}y(r)
\frac{\sin{(lr/l_c)}}{(lr/l_c)}dr,
\label{eq:3dyft}
\end{equation}
where $l_c\equiv d_A/r_c$ and the 
angular diameter distance in a flat $\Lambda$CDM model is
\begin{equation}
d_A(z)=\frac{c}{(1+z)H_0}\int_0^{z}\frac{dz'}
{\sqrt{\Omega_m(1+z')^3+\Omega_{\Lambda}}},
\end{equation}
with $H_0$, $\Omega_m$ and $\Omega_{\Lambda}$ denoting the Hubble 
constant today and matter and vacuum energies in critical 
density units, respectively. Note that the neutrino energy density is contained 
in $\Omega_m$ since all neutrino masses considered in this work are certainly 
non-relativistic out to the highest-redshift clusters considered in this work, 
$z\approx 1$. The virial radius of the cluster is related 
to its virial mass via the relation $M_v=\frac{4\pi}{3}R_v^3\Delta_v(z)$, 
where $\Delta_v(z)$ is the overdensity at virialization with respect 
to the background, calculated from the spherical collapse model in 
$\Lambda$CDM universe. 

An expression for the virial radius, $R_{v}$, as a function of the virial mass, 
$M_{v}$, and the underlying cosmology can be derived noting that
\begin{eqnarray}
M_v=\frac{4\pi}{3}R_v^3\rho_v(z) 
=\frac{4\pi}{3}R_v^3\Delta_v(z)\frac{\rho_c(z)}{\rho_c(0)}\rho_c(0),
\end{eqnarray}
where $\rho_c(z)$ and $\Delta_c(z)\equiv\rho_{\nu}(z)/\rho_{c}(z)$ denote 
the critical density and the virial overdensity at redshift $z$, respectively. 
This relation leads to 
\begin{eqnarray}
R_{v}=\left[\frac{3M_{v}}{4\pi\Delta_{c}(z)}\frac{\rho_{c}(0)}{\rho_{c}(z)}
\frac{1}{\rho_{c}(0)}\right]^{1/3}
=1.69\left[\frac{M}{M_{15}}\frac{18\pi^2}{\Delta_{c}(z)}\right]^{1/3}
\left[\frac{\rho_{c}(0)}{\rho_{c}(z)}\right]^{1/3}\,Mpc,
\end{eqnarray}
where $M_{15}$ is the cluster mass in units of $10^{15}M_{\odot}$. 
Writing 
\begin{eqnarray}
\frac{\rho_c(0)}{\rho_c(z)}=\frac{H_0^2}{H(z)^2},
\end{eqnarray}
and the explicit redshift dependence of the Hubble parameter 
\begin{equation}
H^2(z)=H_0^2\left[\Omega_{\Lambda}+\Omega_m(1+z)^3\right],
\end{equation}
we obtain
\begin{eqnarray}
\frac{\rho_c(0)}{\rho_c(z)}={}&\frac{1}{\Omega_{\Lambda}+\Omega_m(1+z)^3}
=\frac{\Omega_m(1+z)^3}{\Omega_{\Lambda}+\Omega_m(1+z)^3}\frac{1}{\Omega_m(1+z)^3}
\equiv\frac{\Omega_m(z)}{\Omega_m}\frac{1}{(1+z)^3}. 
\end{eqnarray}
Substituting this relation in the expression for the virial radius leads to 
\begin{equation}
R_v=1.69\left[\frac{M}{M_{15}}\frac{18\pi^2}{\Delta_c(z)}\right]^{1/3}
\left[\frac{\Omega_m(z)}{\Omega_m}\right]^{1/3}\frac{1}{1+z}\,Mpc
\cdot h^{-1}.
\end{equation}

Hot intracluster (IC) gas is assumed to follow a $\beta$ density profile, 
\begin{eqnarray}
n_{e}(r)=n_{e,0}[1+(r/r_{c})^{2}]^{-3\beta/2}
\end{eqnarray} 
with the observationally-deduced $\beta=2/3$, and a core radius defined as 
$10 r_c=R_v$. The central electron gas density is 
$n_{e,0}=\rho_{g,0}(X+1)/(2 m_p)$, where $X$ and $m_p$ denote the Hydrogen 
mass fraction (taken to be 0.69) and the proton mass, whereas
\begin{equation}
\rho_{g,0}=\frac{f_{g}\cdot M_v}{4\pi r_c^3\int_0^{10}x^2(1+x^2)^{-3\beta/2}dx} 
\end{equation}
is the central gas density, assuming that IC gas constitutes a fraction 
$f_{g}=0.12$ of the total virial mass. The virial expression is adopted for the 
gas temperature 
\begin{equation}
k_BT_{e,0}=\frac{\mu m_p G}{3\beta}\frac{M_v}{R_v},
\end{equation}
valid for $R_v\gg r_c$, where $\mu=0.59$ is the mean molecular weight.
The resulting dependences of the core radius, gas density and temperature on the 
cluster mass and redshift obviously stem from these definitions.

A key quantity employed in the next section in the calculation 
of the signal-to-noise detection of an individual cluster of mass $M$ at 
redshift $z$ is the power spectrum generated by the cluster
\begin{eqnarray}
C_{l}^{SZ}(M,z;M_{\nu})=\frac{1}{4\pi}|\tilde{\xi}(l,M_{\nu};M,z)|^{2}\mathcal{F}^{2}(x).
\end{eqnarray}
Clearly, the total contribution from the entire cluster population to the SZ 
signal (which is irrelevant to this work) is obtained from Eq.(23) by a weighted 
integration over $M$ and $z$ with $\frac{dn}{dM}\frac{dV}{dz}$, i.e. integration over 
differential cluster number $dN$ (Eq.6).
The SZ power spectrum is shown in Fig.1 for the three fiducial neutrino masses 
considered in this work: 0.1 (black), 0.3 (blue), and 0.6 (red) eV. 
All power spectra were obtained by setting $\mathcal{F}=-2$ which 
corresponds to the Rayleigh-Jeans limit. 
The appreciable sensitivity to the assumed neutrino mass comes from both the 
strong dependence of the transfer function on $f_{\nu}$ (Eqs.2, 1 \& 5), i.e. 
$\Delta\sigma_{8}/\sigma_{8}\approx -8\Delta f_{\nu}$ and the scaling of the 
SZ power spectrum with high power of $\sigma_{8}$, typically $\sigma_{8}^{7}$.

The contribution to the SZ power spectrum from cluster correlations is 
typically an order of magnitude lower than the Poissonian component, 
and peaks on angular degree scales where it is overwhelmed by the primary 
CMB anisotropy. Consequently, its expected relative contribution to the 
angular power spectrum is small (e.g., [40, 41]) and we therefore do not 
consider it in our analysis. Also ignored is the small kinematic component 
of the SZ effect [36]. Had we included these two components in our analysis, 
cluster detection rate would have increased somewhat, improving our 
statistics and therefore lowering the uncertainty on neutrino masses, 
but these effects are very small.

\section{Fisher Matrix Analysis}

Our cosmological model includes the normalization $A$ and tilt $n$
of the primordial scalar perturbations, neutrino, $\Omega_{\nu}h^{2}$, 
$\Omega_{m}h^{2}$, and baryon, $\Omega_{b}h^{2}$, energy densities, the 
Hubble parameter $h$ in km/sec/Mpc units, 
dark energy equation of state $w$, optical depth for 
reionization $\tau$, and the primordial helium abundance $Y_{p}$.
The last two parameters do not directly affect the SZ quantities 
but are correlated with other cosmological parameters which might affect the 
uncertainty of these and indirectly impact neutrino mass when probed with 
SZ number counts (and power spectrum). We first carry out the forecasts for 
the primary CMB. This is done both with and without lensing extraction 
(LE). The corresponding Fisher matrices are denoted $F^{pr}$ and 
$F^{LE}$, respectively.

The extremely steep dependence of the SZ effect on $\sigma_{8}$ is clearly 
expected to result in strong dependence of neutrino mass constraints on the 
fiducial neutrino mass. This follows from the fact that for lower neutrino masses 
(when $A$ is fixed) $\sigma_{8}$ increases 
and as a result the SZ amplitude increases, boosting the signal-to-noise 
($\mathcal{S/N}$) of cluster detection and tightening the constraints on 
inferred neutrino mass. Clearly, the main cluster parameter which 
determines the magnitude of the SZ signal, and 
thereby whether or not it will be detected in a flux-limited survey, is 
the total mass. However, the cluster redshift is also of relevance, as 
explained above and will be further discussed 
below. Since the mean properties of the cluster population provide a good 
basis for estimating global parameters, their gaseous mass fraction could, 
in principle, be related directly to the ratio $\Omega_{b}/\Omega_{m}$ 
(after accounting for the small contribution by the cluster member galaxies).
However, in this work we adopt the fiducial value of the gas mass fraction deduced 
from the sample of SZ measurements with the BIMA \& OVRO telescopes, 
$f_{g}=0.12$ [42]. Due to the appreciable uncertainty in the measured value 
of $f_{g}$ this cluster parameter is also treated as a model parameter in 
our Fisher matrix analysis.

In calculating the Fisher matrix for the primary CMB (with and without 
LE) we follow the standard approach which we do not reproduce here; 
details of the calculation can be found in (e.g.) [17].

\subsection{$M_{\nu}$ from Cluster Number Counts and Correlations}

As in, e.g., [43], we write for the Poissonian likelihood function 
for cluster number counts in the $i$'th redshift bin 
\begin{eqnarray}
\mathcal{L}_{i}(N_{i}^{p};N_{i}^{o})
\propto\frac{(N_{i}^{p})^{N_{i}^{o}}\exp(-N_{i}^{p})}{N_{i}^{o}!} ,
\end{eqnarray}
where $N_{i}^{p}$ is the predicted 
(for a given cosmological model and specified neutrino mass) and 
$N_{i}^{o}$ the observed cluster number in the $i$'th redshift bin. 
Note that here and in the following we use abbreviated notation and 
denote $\Delta N_{i}$ of Eq.(7) as $N_{i}$. If $N_{i}^{p}$ is a 
function of several cosmological parameters $\lambda_{k}$, we may 
consider small deviations with respect to the expected fiducial value 
$N_{i}^{o}\approx N_{i}^{p}+\sum_{k}\frac{\partial N_{i}^{p}}
{\partial\lambda_{k}} \Delta\lambda_{k}$. The Fisher matrix for cluster 
number counts can then be determined from the likelihood function 
\begin{eqnarray}
F_{jm}^{N}=-\frac{\partial^{2}\mathcal{\ln L}}{\partial\lambda_{j}\partial\lambda_{m}}
=\sum_{i}\frac{1}{N_{i}}\frac{\partial N_{i}}
{\partial\lambda_{j}}\frac{\partial N_{i}}{\partial\lambda_{m}}.
\end{eqnarray}
The estimated uncertainty in the parameter $\lambda_{j}$ is then
\begin{eqnarray}
\Delta\lambda_{j}=(F_{jj}^{N})^{-1/2} ,
\end{eqnarray}
where we take the square root of the $j$'th Fisher matrix element. 

The second neutrino probe we employ in this work are measurements of 
the SZ power spectrum. Again, only those clusters which are detected at 
$5\sigma$ or better qualify for our `simulated' sample. For the corresponding 
Fisher matrix we use the Fourier transform $P_{c}(k;z)$ of the cluster 
correlation function $C(r;z)$, as in [44]
\begin{eqnarray}
F_{jm}^{P_{c}}&=&\sum_{a,i}
\frac{\partial\ln(k_{\perp}^{2}k_{\parallel}P_{c})_{ai}}{\partial\lambda_{j}}
\frac{\partial\ln(k_{\perp}^{2}k_{\parallel}P_{c})_{ai}}{\partial\lambda_{m}}
\frac{(V_{k}V_{eff}(k))_{ai}}{2}
\end{eqnarray}
where the effective survey volume is
\begin{eqnarray}
V_{eff}(k)=4\pi f_{sky}\int r^{2}(z)\frac{dr}{dz}dz
\left[\frac{n(z_{i})P_{c}(k)}{1+n(z_{i})P_{c}(k)}\right]^{2}
\end{eqnarray}
and $n(z_{i})$ is the number density of detected clusters at the i'th redshift bin.
$V_{k}$ is the k-space volume element
\begin{eqnarray}
V_{k}=\frac{2\pi\Delta(k_{\perp}^{2})\Delta k_{\parallel}}{(2\pi)^{3}}.
\end{eqnarray}
$P_{c}(k_{\perp},k_{\parallel};z)$, the cluster power spectrum, is
\begin{eqnarray}
P_{c}(k_{\perp},k_{\parallel};z)=[1+\beta(z)\mu^{2}]^{2}b^{2}(z)P(k;z)
\end{eqnarray}
where $k^{2}=k_{\perp}^{2}+k_{\parallel}^{2}$, $\mu\equiv k_{\parallel}/k$, 
$\beta(z)\equiv\frac{1}{b(z)}\frac{d(\ln D_{gr})}{d(\ln a)}$ 
and $D_{gr}$ is the linear growth factor obtained from solving the equation 
governing the linear evolution of density perturbations $\delta$
\begin{eqnarray}
\frac{d^2\delta}{da^2}+\left[\frac{2}{a}-\frac{1}{2a}
\left(\frac{\Omega_m a^{-3}-2\Omega_{\Lambda}}{\Omega_m a^{-3}
+\Omega_{\Lambda}}\right)\right]\frac{\delta}{da}-\frac{3}{2}
\frac{\Omega_m(1-f_{\nu})}{\Omega_m a^{2}+\Omega_{\Lambda}a^{5}}\delta=0
\end{eqnarray}
where $a$ is the scale factor, and $f_{\nu}\equiv\Omega_{\nu}/\Omega_{m}$ 
is the fraction of DM in the form of neutrinos. 
Here, the mass-averaged linear bias is
\begin{eqnarray}
b(z)=\int dM\frac{dn(M,z)}{dM}\tilde{b}(M;z)/\int dM \frac{dn(M,z)}{dM}
\end{eqnarray}
where $\tilde{b}(M;z)$ is obtained from hydrodynamical simulations [45].

The total Fisher matrix that includes cluster number counts combined 
with either the primordial or lensed CMB is $F_{jm}^{pr}+F_{jm}^{N}$ or 
$F_{jm}^{{\rm LE}}+F_{jm}^{N}$, respectively, and the uncertainty in the 
parameter $\lambda_{j}$ is similar to that in Eq.(26), except for the 
fact that now $F_{jm}^{N}$ is replaced with either 
$F_{jm}^{pr}+F_{jm}^{N}$ or $F_{jm}^{{\rm LE}}+F_{jm}^{N}$. Similarly, 
from the power spectrum we obtain $F_{jm}^{pr}+F_{jm}^{c}$ or 
$F_{jm}^{{\rm LE}}+F_{jm}^{c}$, respectively. 

Integrated cluster quantities, such as number counts and the power 
spectrum, obviously depend on the assumed mass range. Our nominal mass 
range is taken to be $10^{13}-3\times 10^{15}$ $M_{\odot}$. While it 
is doubtful whether our description of IC gas properties is valid at 
the low mass end of this range, we have verified that our results are 
insensitive to the exact value of the low mass end by repeating the 
calculations with three different values for the low mass cutoff, with 
only a negligible fraction ($~2\%$ and $~0.2\%$ for EPIC and SPT, 
respectively) of the clusters in our mass range have masses smaller 
than $5\times 10^{13}h^{-1}M_{\odot}$. This is an important 
consistency check since the assumption of virialization (Eqs. 14-22) 
is expected to break down at around this value, and since the cluster 
mass cannot be deduced in a model-independent way, it is reassuring 
that the detection threshold is set such that clusters for which our 
model breaks down are not included in the sample. This essentially is 
the case as long as the detection threshold is taken to be at the 
$5\sigma$ significance level. The upper mass limit of this range is in 
accord with recent detailed joint data analyses (e.g. [46]) that have resulted 
in reliable mass estimates that are already very close to the above value. 
Here too, raising the upper mass to $10^{16}$ $M_{\odot}$ has little impact 
on the uncertainty in inferred $M_{\nu}$.  

To estimate the $\mathcal{S/N}$ with which a cluster can 
be detected in a survey, we assume that the main noise sources are the 
instrument, primary CMB anisotropy, and point source emission.
We ignore other astrophysical foregrounds, such as free-free emission, 
synchrotron radiation and dust, due to their Galactic origin and the  
planned removal of this sky region in the PLANCK and EPIC data analysis. 
Galactic contamination in the SPT and ACT survey is minimized by selecting a 
radio-quiet sky region. Optimal (minimum variance) {\it unbiased} filters 
can be constructed for cluster detection with the resulting $\mathcal{S/N}$ 
variance [47, 48] characterized by minimum confusion noise 
\begin{eqnarray} 
\left(\mathcal{\frac{S}{N}}\right)^{2}=
\sum_{i}\int\frac{ldl}{2\pi}\frac{|\tilde{\xi}_{l}|^{2}(\nu_{i})}
{[C_{l}^{det}(\nu_{i})+C_{l}^{CMB}+C_{l}^{PS}(\nu_{i})]} , 
\end{eqnarray}
which is merely proportional to the ratio of the power spectra of 
signal and noise weighted by 
Fourier space volume ($\propto l$) and summed over all accessible multipoles. 
Contribution to the $\mathcal{S/N}$ from scales below the beamsize are 
exponentially damped 
\begin{eqnarray} 
C_{l}^{det}=(\Delta_{T}\theta_{b})^{2}e^{l^{2}\sigma_{b}^{2}}
\end{eqnarray}
where $\theta_{b}=\sqrt{8\ln(2)}\sigma_{b}$. $\Delta_{T}$ is a 
noise-equivalent-temperature (NET) of the experiment, and the sum over $i$ 
runs over all the detector frequency channels.
The relevant parameters for the PLANCK instrument, SPT, and EPIC are 
listed in Table I. The adopted power spectrum of point sources is [49]
\begin{eqnarray}
C_{l}^{PS}=A_{ps}(x_{Q}/x)^{-4}(e^{x}-1)^{4}x^{-4}e^{-2x} ,
\end{eqnarray}
where $A_{ps}=0.015\mu K^{2}$-str, $x_{Q}\equiv(h\nu_{Q})/(kT)$, and 
$\nu_{Q}=40.7 GHz$ is WMAP's Q-band. $C_{l}^{PS}$, $\Delta_{T}$, and $\theta_{b}$ 
are all frequency-dependent. 

As shown in Eqs.(12) \& (23) the SZ power spectrum is a function of the cluster 
core radius, $\theta_{c}$, which in itself is a function of $M$ and $z$ (Eqs. 19 \& 20). 
Therefore, one can define the $\mathcal{S/N}$ per a single cluster in the 
$M$-$z$ cell
\begin{eqnarray}
\left(\mathcal{\frac{S}{N}}\right)_{M,z}^{2}\equiv
\sum_{i}\int\frac{ldl}{2\pi}\frac{|\tilde{\xi_{l}}(M,z;\nu_{i})|^{2}}
{[C_{l}^{det}(\nu_{i})+C_{l}^{CMB}+C_{l}^{PS}(\nu_{i})]},
\end{eqnarray}
where $\tilde{\xi_{l}}(M,z;\nu_{i})$ is given by Eq.(12); recall though 
that the cluster core radius is a function of both mass and redshift. 
This measure, together with the mass function, which determines the 
number density in each of the $M$-$z$ cells, can be used to estimate 
how many clusters will be detected at a given cell by a specific 
experiment. This criterion is employed in the next section. 

\section{Results}

In this work we have adopted the $\Lambda$CDM cosmological model with 
WMAP best-fit parameters. The cluster population is described in terms of 
the mass function of Tinker et al. [34, 45], and the (hot) IC gas mass fraction 
was taken at the level deduced from SZ observations, $f_g \simeq 0.12$. IC gas 
has been assumed to be isothermal with a beta density profile. 
It is important to note that in this analysis no priors were set neither on 
the cosmological parameters nor on $f_{g}$. We reiterate that 
changing neutrino masses over the range [0.1, 0.6] eV changes 
$\sigma_{8}$ by $\sim 25\%$. Since the SZ effect strongly depends 
on $\sigma_{8}$, we expect observational measures of the effect to show 
a corresponding level of sensitivity to varying $M_{\nu}$, as is indeed 
evident from Figure 1.

Our analysis includes several inherent noise sources - detector noise, 
and the noise due to the primary CMB anisotropy and extragalactic point 
sources. We first calculate the distribution of $(\mathcal{S/N})_{M,z}$ 
with which clusters of mass $M$ at redshift $z$ will be detected (Eq. 
36) in the PLANCK, SPT and proposed EPIC surveys. As is clear from Eq.(36)
the clusters detectability likelihood depends on both their masses and redshifts; 
the mass and redshift together determine the $\mathcal{S/N}$, as is clear from 
the fact that the gas temperature and core radius (both physical and certainly 
its angular projection on the sky) depend on both $M$ and $z$ (Eqs. 19 \& 
20). We set the level of statistical detectability to $5\sigma$.

Our basic results for neutrino mass uncertainty based on cluster number counts 
and power spectrum from SZ surveys with the ongoing PLANCK, SPT, and 
proposed EPIC projects are presented in Tables II and III for the fiducial 
neutrino masses, $0.1$, $0.3$, and $0.6$ eV. In Table II we show (from left 
to right) the expected uncertainty on the inferred $M_{\nu}$ from the 
primordial CMB (both temperature anisotropy and polarization), primary CMB and 
cluster SZ number counts, primary CMB and cluster correlation, 
and finally the constraints from joint primary CMB, cluster number counts 
and power spectrum. Table III is the same as Table II, but the priors are from 
LE. It is particularly interesting to see that if $M_{\nu}$ is $\sim 0.1 eV$, then 
the joint primordial CMB, cluster abundance and correlation analysis for 
PLANCK yields $\sigma_{M_{\nu}}=0.28 eV$, whereas the constraint from CMB LE 
is $0.14 eV$.

Similar results for ACT are shown in Table V. This project is 
viewed here as a generic ground-based SZ survey experiment with deep exposure 
on a very small patch of the sky. Our goal is to optimize the scanning 
strategy for neutrino mass inference. The analysis presented here 
clearly illustrates that the uncertainty on neutrino masses can be 
reduced by $\sim 1/2$ when the survey area increases from 200${\rm deg^{2}}$ 
to 4,000${\rm deg^{2}}$. Indeed, increasing $f_{sky}$ results in a 
higher noise per sky-pixel, which lowers the detection significance of 
many clusters, but with the $5\sigma$ detection-threshold we adopt here it 
turns out that only the smallest clusters are removed from our sample 
and this is outweighted by improving the statistics of massive/nearby 
clusters by virtue of increasing $f_{sky}$. Specifically, we considered 
four fiducial ACT modes; ACT-A, B, C and D (Table IV). ACT-A refers to the 
nominal ACT experiment that covers 200${\rm deg^{2}}$ of the sky. ACT-B, C 
and D scan 400, 1000 and 4,000 ${\rm deg^{2}}$, respectively. 
The larger $f_{sky}$, the shorter is the integration time available per 
sky-pixel, and assuming the integration is uniform over the sky and that 
the instrumental noise is inversely proportional to the integration time, 
$\Delta_{T}\propto t_{int}^{-1/2}$, we can rescale ACT-A to the other 
three modes. While ACT-A has very low noise it is limited by the lower 
statistics because of the survey area, 0.5\% of the sky. Results listed 
in Tables V and VI clearly illustrate that eventually the gain in sky area 
(in the cases of ACT-C and D) result in a notable improvement in neutrino mass 
constraints. The lesson from this is that observing a larger sky-patch is 
preferable, at least for neutrino mass inference and for a CMB telescope 
with ACT-like specifications. It should be noted here that the priors, i.e. 
constraints that come from either primordial CMB or LE that we 
adopted in these tables come from PLANCK, not ACT itself (which should 
also be affected by changing $f_{sky}$), and are for a fixed $f_{sky}=0.65$.

An important aspect of the theoretical treatment is careful selection of the 
mass range. In principle, we could compare expected and observed number counts 
on the entire 2D $M$-$z$ space, if it was realistic to reliably deduce cluster 
masses (from the `blind' surveys considered here). Since cluster mass inference 
is highly model-dependent we always integrate the cluster distribution over 
mass and consider number counts in redshift-space only, as described earlier 
(Eqs. 25 \& 27). Integrating the {\it expected} number counts and cluster 
correlation over mass we have to impose a low mass cutoff, thereby removing all 
low-mass clusters from our sample. As noted already, this is motivated 
mainly by the fact that objects that have detectable mass of 
sufficiently hot ($\geq 1$ keV) gas, and for which our theoretical 
description is still valid - including aspects such as the gas density 
profile, mass-temperature scaling, and indeed the attainment of 
hydrostatic equilibrium in the first place - do not include very poor 
clusters and groups of galaxies. We have checked and verified that our 
results are robust with respect to changing this low mass cutoff to the 
lowest cluster mass of $10^{13}M_{\odot}$ we consider in this work. This 
obviously reflects the consistency between our theoretically selected value 
of the low mass cutoff and the high significance that we require for the 
threshold for SZ detection. 

Formation of clusters is relatively late in the $\Lambda$CDM model, with 
the vast majority of medium to high mass clusters forming at $z<1$ (as is 
clear from plots of the probability distributions of formation times in [46]). 
We checked that essentially irrespective of the fiducial neutrino mass 
considered, the contribution to the Fisher matrix saturates at 
$z_{max}\lesssim 1$. Higher-redshift clusters are simply not sufficiently 
massive and in addition project on too small angular scales to be detected.

Redshift binning may also introduce some bias; by increasing bin width 
the diagnostic capability is weakened, but the impact of random fluctuations 
e.g. modelling errors that bias the redshift of cluster formation 
is reduced. This motivated us to determine an optimal redshift-binning in 
our number counts analysis by binning the redshift interval 
$[0-1]$ into 10, 20 and 40 linearly spaced redshift shells. We found that 
increasing this number from 10 to 20 improves the statistical error on 
the deduced neutrino mass by not more than $\sim 30\%$. Further binning 
refinement to 40 shells reduces the error by at most $15\%$, implying 
that further increase in the number of shells does not reduce 
the statistical uncertainty but may come at a cost of a higher level of 
systematic error. All results reported in this work using cluster number counts 
were obtained by binning the simulated data into 40 redshift bins. For the 
cluster correlation probe we used only 10 redshift shells; as in [51]
we choose k-binning in Eq.(27) to be wide enough for different bins to 
be uncorrelated. We assume $\Delta z=0.1$ for this analysis and 
$\Delta k=0.005 h Mpc^{-1}$ in the k-range [0.005-0.1]$h Mpc^{-1}$.

The most important source of uncertainty in modeling cluster abundance and 
internal properties is the mass function. Current uncertainties in this 
basic function were explicitly included in our analysis. Continued extensive 
cosmological hydrodynamical simulations (e.g. [52]) are 
likely to result in a significantly more precise description of this function 
across the cluster mass range. In contrast, uncertainties stemming from using 
simple models for the spatial profiles of the gas density and temperature are 
of secondary importance, simply because these are much reduced when calculating 
integrated SZ measures and, more generally, will have little effect on cluster 
detection. After all, the magnitude of the effect in a cluster is 
not determined by each of these quantities separately (in the non-relativistic 
limit that is valid for our purpose here), but rather by the integrated gas 
pressure along the line of sight. Thus, even though central regions of clusters 
are roughly isothermal, and the gas density profile is well approximated by 
the beta profile, realistic deviations from isothermal beta profile for the 
pressure, when integrated across a cluster and over the mass function, introduce 
a level of uncertainty that is ignorable for our purposes here. 
Mass function normalization and three additional shape parameters (Eq. 9) 
are included in our analysis.

\section{Discussion}

The CMB is perhaps one of best understood and model-independent 
underpinnings of modern cosmology. Temperature anisotropy and 
polarization measurements of the primary CMB signal already taught 
us a great deal about the geometry of the universe, its energy content, 
reionization history, etc. With the imminent detection of the 
lensing-induced signal of the CMB by PLANCK, and possibly even the 
inflationary-induced B-mode polarization (if inflation is indeed related 
to GUT scale physics), the CMB may open new windows into the physics of 
neutrinos as well as processes that occurred in the GUT era; the latter are 
well beyond the reach of any present or future terrestrial experiment. 

However, the capacity of primary CMB alone to constrain $\sim Mpc$ 
cosmology is rather limited; even CMB lensing by the LSS, a sensitive 
probe of neutrino masses, takes place on considerably larger physical 
scales. In addition, standard forecasts of LE performance widely build on the 
assumption that the CMB signal is gaussian in the absence of lensing, 
and any nongaussianity present in the data can be attributed to lensing 
of the CMB. It should be emphasized that the results for neutrino masses 
derived from CMB and LE reported here, and elsewhere in the literature, 
always make this simplifying assumption. However, non-gaussianity can also 
be induced by astrophysical sources. The issue of primordial nongaussianity 
is still open and should be considered as another source of confusion. 
On the relevant angular scales secondary CMB signals, such as the SZ effect, 
are known to be present. These are typically nongaussian and can interfere 
with LE.

Structure on $Mpc$ scales probes the entire evolution history  
of matter perturbations down to these scales. This is especially relevant 
to neutrino physics via the effect of neutrino free streaming on these 
and larger scales. Indeed, $Ly\alpha$ observations can supplement the CMB 
on these scales but since the astrophysics of the former is more involved
it is desirable to consider other probes of 
clustering on Mpc scales to complement $Ly\alpha$, as well as other 
traditional probes such as galaxy clustering, shear measurements, etc., 
or at the very least to serve as a consistency check. 
The possibility of using cluster catalogues in constraining neutrino 
masses has already been discussed in [19]. More generally, extraction 
of cosmological parameters, such as $\Omega_{b}$, $\Omega_{\Lambda}$, from 
cluster number counts - in conjugation with other cosmological probes - 
is widely discussed in the literature, e.g. [43,53]. 

In this work we considered the prospect of SZ surveys to constrain 
neutrino masses. We have shown that using cluster number abundance alone 
neutrino mass uncertainties may be constrained to the $\sim 0.36-0.55$ eV 
levels by PLANCK and $\sim 2-3$ times weaker by SPT SZ surveys alone, on a 
wide range of allowed neutrino masses ($M_{\nu}<0.6$ eV). The proposed 
EPIC satellite with its SZ survey of thousands of clusters will be able 
to set impressive constraints on neutrino masses at the level that will 
enable us to constrain models of inverted mass hierarchy (where $M_{\nu}>0.1$ 
eV). For this seemingly promising potential to be realized a full account of 
systematics has to be achieved; we elaborate on mass-function uncertainties 
below. Systematics, such as unknown gas fraction $f_{g}$, 
are easily accounted for by simply including them as free parameters in our 
analysis; with predicted hundreds (SPT) or thousands (PLANCK and EPIC) 
clusters we have sufficient freedom to include other parameters, such 
as $f_{g}$, in addition to $M_{\nu}$ and $A$. It is quite likely that 
follow-up X-ray observations will yield a more precisely determined value of 
$f_{g}$, a possibility we clearly do not consider in this work as we assumed 
no prior on $f_{g}$. 

Another subtlety comes from possible degeneracy with $\Omega_{de}$. 
Including degeneracy with dark energy in our framework is indeed possible 
with the code employed here for the transfer function [30] as it was 
developed especially for the purpose of calculating the impact of large scale 
structure suppression by both neutrino free streaming and accelerated expansion 
of the universe induced by dark energy. A very crude estimate of this effect can 
be obtained by fixing the CDM component, $\Omega_{dm}$, and changing 
$\Omega_{\nu}$ on the expense of $\Omega_{de}$, 
but since the total energy density is assumed to be fixed to its critical value, 
and since both neutrino (with a total mass in the range considered here) 
and dark energy do not clump on galaxy cluster scales, we can conclude 
that - at least to first approximation - allowing all three $\Omega_{\nu}$, 
$\Omega_{de}$ and $\Omega_{dm}$ to vary (subject to the constraint that the 
total energy density is fixed) does not substantially affect our results.

The impact of sub-eV neutrino masses will obviously be reflected also in the 
superposed SZ power from the population of clusters. We have calculated the 
full power spectrum by integrating the expression for a single cluster (specified 
in Section III) over the mass function for three neutrino masses, $0.1$, $0.3$, and 
$0.6$ eV. Results of this calculation are shown in Fig. 1 by the black, blue, and 
red curves, respectively. Since $\sigma_{8}$ is approximately proportional to 
$1-8f_{\nu}$, changing neutrino mass in the range $[0.1-0.6]$ eV corresponds to 
varying $f_{\nu}$ in the range $[0.7-4.3]\times 10^{-2}$; this introduces a 
fractional change $\Delta\sigma_{8}/\sigma_{8}\approx -8\Delta f_{\nu}$. 
The steep dependence of SZ power on $\sigma_{8}$ translates to a relative change 
of $\approx 56f_{\nu}$. From Fig. 1 we see that this factor amounts to an overall 
variation of the peak power by a factor of $\sim 3$ for the above range of neutrino 
masses. Note though that in our calculations we used the 
precise transfer function and did not employ this crude approximation which 
we mention here only for the sake of obtaining a rough estimate of the SZ 
sensitivity to changing neutrino masses. Confusion 
with the CMB primary and other confusing signals renders this neutrino signature 
less diagnostically useful as compared with the more optimized cluster SZ number 
counts or correlations. It is important to note that the steep SZ 
power spectrum dependence on neutrino masses does not {\it directly} imply 
similar sensitivity in number counts. The only direct effect of the power 
spectrum dependence on $M_{\nu}$ is that large neutrino mass decreases the 
SZ power spectrum, and therefore the $\mathcal{S/N}$ with which galaxy 
clusters are detected, thereby boosting the uncertainty on neutrino masses 
as is illustrated in Table II.

Another tantalizing possibility is using these large cluster surveys to {\it 
simultaneously} calibrate the mass function and determine neutrino mass. This 
will help reducing possible bias that could be caused by miss-calibration of 
the mass function. 
In our analysis we adopted the functional form of the Tinker mass function, 
Eqs.(8) \& (9), and considered the shape parameters as free parameters in 
the statistical analysis, adopting the central values reported in [34]. 
To test the robustness of our results to changing the mass function used 
in the parameter estimation we repeated the analysis with a different mass 
function which was similarly derived from hydrodynamical simulation - the 
Jenkins mass function [54]. We found that the constraints on neutrino mass 
do not change by more than 4\%, and in most cases the differences do not 
exceed the 1-2\% level. 

It is clear that more can be done to refine the analysis presented here since, 
admittedly, cluster physics and statistical properties of the cluster population 
are model-dependent. However, we believe that the adoption of a rigorous criterion 
for cluster detection in $M$-$z$ space, assuming implementation of well-known 
unbiased optimal filters (in conjugation with the need to reject low mass 
clusters from our sample), the realization that the statistics improve when 
neutrino masses are lower as $\sigma_{8}$ increases, the use of a high precision 
transfer function for neutrinos in our theoretical SZ model, and optimal binning 
in redshift space, all these factors together constitute important 
steps towards achieving a high precision probe of neutrino masses. These 
features lend credibility to the results presented in this work. 
The results are encouraging: Cluster number counts, 
and to a lesser extent cluster correlations, may indeed be reliable neutrino 
probes if carefully analyzed. The utility of this probe largely depends on the 
neutrino mass; the lower the mass, the tighter is the constraint (as is shown 
in Table II and explained above). The prospects will only be further enhanced 
when SZ surveys will be jointly analyzed with results from extensive X-ray surveys.

\ \\
\ \\
\ \\


\begin{figure*}[h]
\centering
\epsfig{file=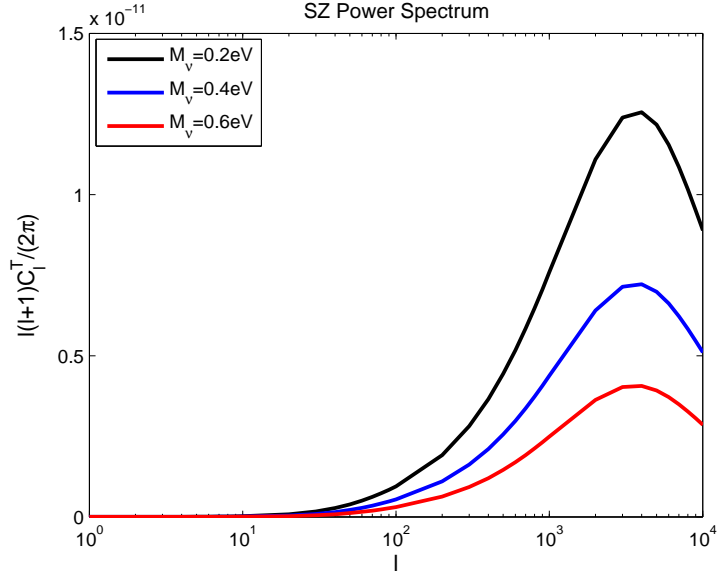, width=10.5cm, clip=}
\caption{SZ power spectrum for three neutrino masses: 0.1 (black), 0.3 (blue) and 0.6 
eV (red). As explained in the text, the dependence $\sigma_{8} \propto 1-8f_{\nu}$ 
translates to a relative change of $\approx 56f_{\nu}$ in the value of $\sigma_{8}$ 
when the total neutrino mass assumes values in the range $0.1-0.6$ eV.}
\end{figure*}

\begin{table}
\begin{tabular}{|c|c|c|c|c|}
\hline
Experiment & $f_{\rm sky}$ & $\nu [GHz]$ & $\theta_b [1']$ & $\Delta_T [\mu K]$\\
\hline
\hline
& &  30 & 33 &  4.4\\
& &  44 & 23 &  6.5\\
& &  70 & 14 &  9.8\\
& &  100 & 9.5 &  6.8\\
PLANCK&0.65& 143 & 7.1 & 6.0\\
& &  217 & 5.0 &  13.1\\
&   & 353 & 5.0 & 40.1\\
& &  545 & 5.0 &  401\\
&   & 856 & 5.0 & 18300\\
\hline
&& 150 & 1.0 & 13.4\\
SPT &0.10& 274 & 0.52 & 71.4\\
& & 345 & 0.29 & 583.9\\
\hline
& &  30 & 28 &  0.50\\
& &  45 & 19 &  0.30\\
& &  70 & 12 &  0.21\\
&& 100 & 8.4 & 0.22\\
EPIC (4K)&0.65& 150 & 5.6 & 0.25\\
& & 220 & 3.8 & 0.66\\
& & 340 & 2.5 & 2.24\\
& & 500 & 1.7 & 9.41\\
& & 850 & 1.0 & 740.0\\
\hline
\end{tabular}
\caption{Sensitivity parameters for Planck, SPT, and EPIC: $f_{\rm sky}$ is the 
observed fraction of the sky, $\nu_0$ is the central channel frequency (in GHz), 
$\theta_b$ is the FWHM (Full-Width at Half-Maximum) in arcminutes, and $\Delta_{T}$ 
is the temperature sensitivity per pixel in $\mu$K. Polarization sensitivity is
$\Delta_{E}=\Delta_{B}=\sqrt{2}\Delta_{T}$, except for the 545 and 857 GHz bands of 
PLANCK which are unpolarized. We assume all the frequency bands can be used for 
primordial CMB and LE, but for the purpose of cluster detection we assumed that 
only the 100, 143 \& 353 GHz (PLANCK) and 100, 150 \& 340 GHz channels (EPIC) 
can be reliably used, since these are the least foreground-contaminated bands.}
\end{table}

\begin{table}
\begin{tabular}{|c|c|c|c|c|c|}
\hline
Experiment&$M_{\nu}[eV]$&$\sigma_{M_{\nu}}[eV]$(prim.)&$\sigma_{M_{\nu}}[eV]$[prim.+N(z)]&$\sigma_{M_{\nu}}[eV]$[prim.+Pc(z)]&
$\sigma_{M_{\nu}}[eV]$[prim.+N(z)+Pc(z)]\\
\hline
 &0.1&&0.46&1.21&0.44\\
SPT&0.3&2.02&1.25&1.37&1.01\\
 &0.6&&1.52&1.55&1.27\\
\hline\hline
 &0.1&&0.36&0.55&0.28\\
PLANCK&0.3&0.67&0.55&0.60&0.42\\
 &0.6&&0.55&0.63&0.50\\
\hline\hline
 &0.1&&0.13&0.23&0.12\\
EPIC&0.3&0.34&0.25&0.26&0.20\\
 &0.6&&0.30&0.30&0.25\\
\hline\hline
\end{tabular}
\caption{Statistical uncertainty on total neutrino mass from cluster number counts 
obtained from PLANCK, SPT, and EPIC SZ surveys (no LE). Shown are neutrino mass uncertainties 
($1\sigma$) expected to be obtained from the respective planned surveys. We consider 
$\sigma_{M_{\nu}}$ assuming fiducial masses 0.1, 0.3 and 0.6 eV. We show the results from CMB alone, 
CMB+cluster number counts N(z), CMB+cluster correlation Pc, and CMB+N(z)+Pc.}
\end{table}

\begin{table}
\begin{tabular}{|c|c|c|c|c|c|}
\hline
Experiment&$M_{\nu}[eV]$&$\sigma_{M_{\nu}}[eV]$(LE)&$\sigma_{M_{\nu}}[eV]$[LE+N(z)]&$\sigma_{M_{\nu}}[eV]$[LE+Pc(z)]&
$\sigma_{M_{\nu}}[eV]$[LE+N(z)+Pc(z)]\\
\hline
 &0.1&&0.35&0.41&0.25\\
SPT&0.3&0.52&0.50&0.44&0.34\\
 &0.6&&0.51&0.49&0.43\\
\hline\hline
 &0.1&&0.12&0.12&0.11\\
PLANCK&0.3&0.14&0.12&0.13&0.12\\
 &0.6&&0.13&0.14&0.13\\
\hline\hline
 &0.1&&0.04&0.04&0.04\\
EPIC&0.3&0.05&0.04&0.04&0.04\\
 &0.6&&0.04&0.04&0.04\\
\hline\hline
\end{tabular}
\caption{Same as in Table II but with LE priors.}
\end{table}

\begin{table}
\begin{tabular}{|c|c|c|c|c|}
\hline
Experiment & ${\rm deg^{2}}$ & $\nu [GHz]$ & $\theta_b [1']$ & $\Delta_T [\mu K]$\\
\hline
\hline
ACT-A&200& 145 & 1.7 & 2.0\\
& &  280 & 0.93 &  8.8\\
\hline
ACT-B&400& 145 & 1.7 & 2.83\\
& &  280 & 0.93 &  12.45\\
\hline
ACT-C&1,000& 145 & 1.7 & 4.47\\
& &  280 & 0.93 &  19.68\\
\hline
ACT-D&4,000& 145 & 1.7 & 8.94\\
& &  280 & 0.93 &  39.35\\
\hline
\hline
\end{tabular}
\caption{ACT-A, B, C and D: ACT-A is rescaled to B, C and D by assuming 
$f_{sky}\propto t_{int}^{-1}$ and $\Delta_{T}\propto t_{int}^{-1/2}$.}
\end{table}

\begin{table}
\begin{tabular}{|c|c|c|c|c|}
\hline
Experiment&$\sigma_{M_{\nu}}[eV]$(prim.)&$\sigma_{M_{\nu}}[eV]$[prim.+N(z)]&$\sigma_{M_{\nu}}[eV]$[prim.+Pc(z)]&
$\sigma_{M_{\nu}}[eV]$[prim.+N(z)+Pc(z)]\\
\hline
ACT-A&0.67&0.61&0.66&0.61\\
\hline
ACT-B&0.67&0.60&0.65&0.59\\
\hline
ACT-C&0.67&0.51&0.64&0.49\\
\hline
ACT-D&0.67&0.40&0.61&0.36\\
\hline\hline
\end{tabular}
\caption{Statistical uncertainty on total neutrino mass from cluster number counts 
and cluster correlation (similar to Table II)
obtained from ACT-A, B, C and D SZ surveys (no LE). Shown are neutrino mass uncertainties 
($1\sigma$) expected to be obtained from the respective planned surveys. We consider 
$\sigma_{M_{\nu}}$ assuming fiducial mass 0.1 eV. We show the results from CMB alone, 
CMB+cluster number counts N(z), CMB+cluster correlation Pc, and CMB+N(z)+Pc. 
Primordial CMB priors from PLANCK are considered.}
\end{table}

\begin{table}
\begin{tabular}{|c|c|c|c|c|}
\hline
Experiment&$\sigma_{M_{\nu}}[eV]$(LE)&$\sigma_{M_{\nu}}[eV]$[LE+N(z)]&$\sigma_{M_{\nu}}[eV]$[LE+Pc(z)]&
$\sigma_{M_{\nu}}[eV]$[LE+N(z)+Pc(z)]\\
\hline
ACT-A&0.14&0.14&0.14&0.14\\
\hline
ACT-B&0.14&0.14&0.14&0.14\\
\hline
ACT-C&0.14&0.14&0.14&0.13\\
\hline
ACT-D&0.14&0.13&0.13&0.12\\
\hline\hline
\end{tabular}
\caption{Same as in Table V but with LE priors.}
\end{table}

\end{document}